\documentclass{webofc}
\usepackage[varg]{txfonts}   
\newcommand{\z}{&&\hspace*{-1.5cm}}

\newcommand{\bea}{\begin{eqnarray}}
\newcommand{\eea}{\end{eqnarray}}
\newcommand{\be}{\begin{equation}}
\newcommand{\ee}{\end{equation}}

\begin{document}
\title{EMC effect at small Bjorken $x$ values}
%
%

\author{\firstname{A.} \lastname{Kotikov}\inst{1,2}\fnsep\thanks{\email{kotikov@theor.jinr.ru}}
  \and
        \firstname{B.} \lastname{Shaikhatdenov}\inst{2}
        \and
       \firstname{P.} \lastname{Zhang}\inst{1,3}
}

\institute{
  Institute of Modern Physics, Lanzhou 730000, China
 \and
Joint Institute for Nuclear Research, 141980 Dubna, Russia
\and
University of Chinese Academy of Sciences, Yuquanlu 19A, Beijing 100049, China
          }

\abstract{%
The Bessel-inspired behavior of parton densities at small Bjorken $x$ values is used
along with ``frozen'' and analytic modifications of the
strong coupling constant~\cite{Kotikov:2017mhk} to study the so-called EMC effect.
Among other results, this approach allowed predicting small $x$ behavior of the gluon density in nuclei.
}
\maketitle

\section{Introduction}

The study of deep-inelastic scattering (DIS) of leptons off nuclei reveals
an appearance of a significant nuclear effect that
rules out the naive picture of a nucleus as being a system of quasi-free nucleons
(for a review see, e.g.,~\cite{Arneodo:1992wf}).
It was first observed by the European Muon Collaboration~\cite{Aubert:1983xm} in the valence quark dominance region.

There are two conventional approaches in the field to studying the EMC effect. In the first one, which is at present
more widespread, nuclear parton distribution functions (nPDFs) are extracted from the global fits to nuclear
data by using empirical parametrizations of their normalizations (see~\cite{Eskola:2009uj,Eskola:2016oht})
and the numerical solution to Dokshitzer-Gribov-Lipatov-Altarelli-Parisi (DGLAP)
equations~\cite{DGLAP}
\footnote{Sometimes, in the analyses of DIS experimental data it is convenient to use an exact
  solution to DGLAP equations in the Mellin moment space and reconstruct SF $F_2$ from the moments
 (see recent paper~\cite{Kotikov:2016ljf} and references and discussions therein). The studies of
  nuclear effects in such a type of analyses can be found in~\cite{Krivokhizhin:2005pt}, though its
  consideration is beyond the scope of the present study.}.
The second approach heavily relies on different models of nuclear PDFs~\cite{Kulagin:2004ie}-\cite{Close:1984zn}
(see also recent review~\cite{Kulagin:2016fzf}).

Here we will follow the classic rescaling model~\cite{Jaffe:1983zw,Close:1984zn}, which
is based upon suggestion~\cite{Close:1983tn} that the effective confinement size of gluons and quarks in a
nucleus is greater than in a free nucleon. In the framework of perturbative QCD it was found \cite{Jaffe:1983zw,Close:1984zn,Close:1983tn}
that such a change in the confinement scale predicts that nPDFs and usual (nucleon) PDFs be related by simply rescaling their arguments
(see Eq.~(\ref{va.1a}) below). That is why it is relatively safe to say that the rescaling model
inhales the features of both above approaches: in its framework there
are certain relations between usual and nuclear PDFs that result from shifting the values of kinematical variable $\mu^2$;
however, both densities obey DGLAP equations.

At first, the model was established for the valence quark dominance region $0.2 \leq x \leq 0.8$.
Recently  the range of the model applicability
was extended \cite{Kotikov:2017mhk}
to the small $x$ region,
where the rescaling values can be different for gluons and quarks. To do it
we used the generalized double-scaling approach (gDAS)~\cite{Munich,Q2evo}. The latter is
based upon the analytical solution to DGLAP equations in the small $x$ region and generalizes
earlier studies~\cite{BF1}.

\section{SF $F_2$ at low $x$}

A reasonable agreement between HERA data~\cite{H1ZEUS} and predictions made by perturbative QCD
and observed for $Q^2 \geq 2$ GeV$^2$~\cite{CoDeRo},
confirmed expectations that perturbative QCD is capable of describing the
evolution of parton densities down to very low $Q^2$ values.

Some time ago ZEUS and H1 Collaborations have presented new precise combined data~\cite{Aaron:2009aa}
on the structure function (SF) $F_2$.
An application of the gDAS approach~\cite{Q2evo}
shows that theoretical predictions are well compatible with experimental data at $Q^2 \geq 3\div 4$ GeV$^2$
(see recent results in~\cite{Kotikov:2012sm}).

In~\cite{Kotikov:2017mhk}
we performed the LO analyses of the combined data~\cite{Aaron:2009aa}
for the SF $F_2$, which has the following form
\begin{eqnarray}
  F_2(x,\mu^2) = e \,
  f_q(x,\mu^2),~~~ e=(\sum_1^f e_i^2)/f \, ,
\label{8a}
 \end{eqnarray}
where $e$
is an average of the squared quark charges and
$f_q(x,\mu^2)$ is the
singlet part of quark parton density. The contribution of the nonsinglet part is negligible at low $x$
and therefore omitted in the investigations performed in~\cite{Kotikov:2017mhk}.

We note
that the approach used in these analyses is analogous to that exploited and carried out in NLO ones
in~\cite{Kotikov:2012sm}--\cite{Cvetic1}.
The small-$x$ asymptotic expressions for sea quark and gluon densities $f_a$  can be written as follows
\begin{eqnarray}
\z f_a(x,\mu^2) =
f_a^{+}(x,\mu^2) + f_a^{-}(x,\mu^2),~~~(\mbox{hereafter}~~~a=q,g) \nonumber \\
\z	f^{+}_g(x,\mu^2) = \biggl(A_g + \frac{4}{9} A_q \biggl)
		\tilde{I}_0(\sigma) \; e^{-\overline d_{+} s} + O(\rho),~~
f^{+}_q(x,\mu^2) =
\frac{f}{9} \biggl(A_g + \frac{4}{9} A_q \biggl) \rho \tilde{I}_1(\sigma)  \; e^{-\overline d_{+} s}
+ O(\rho),
\nonumber \\
\z        f^{-}_g(x,\mu^2) = -\frac{4}{9} A_q e^{- d_{-} s} \, + \, O(x),~~
	f^{-}_q(x,\mu^2) ~=~  A_q e^{-d_{-}(1) s} \, + \, O(x),
	\label{8.02}
\end{eqnarray}
where $I_{\nu}$ ($\nu=0,1$)
are the modified Bessel functions
with
\be
s=\ln \left( \frac{a_s(\mu^2_0)}{a_s(\mu^2)} \right),~~
a_s(\mu^2) \equiv \frac{\alpha_s(\mu^2)}{4\pi} = \frac{1}{\beta_0\ln(\mu^2/\Lambda^2_{\rm LO})},~~
\sigma = 2\sqrt{\left|\hat{d}_+\right| s
  \ln \left( \frac{1}{x} \right)}  \ ,~~~ \rho=\frac{\sigma}{2\ln(1/x)},
\label{intro:1a}
\ee
and
\begin{equation}
\hat{d}_+ = - \frac{12}{\beta_0},~~~
\overline d_{+} = 1 + \frac{20f}{27\beta_0},~~~
d_{-} = \frac{16f}{27\beta_0} \, .
\label{intro:1b}
\end{equation}

By using the above results we have analyzed in~\cite{Kotikov:2017mhk} H1 and ZEUS data for $F_2$ \cite{Aaron:2009aa}.
We found (see
Table 1 in \cite{Kotikov:2017mhk}) that the twist-two approximation looks reasonable for $Q^2 \geq 3.5$ GeV$^2$.
It is almost completely compatible with NLO analyses done in~\cite{Kotikov:2012sm}--\cite{Cvetic1}.
Moreover, these results are rather close to original analyses
(see~\cite{Cooper-Sarkar:2016foi} and references therein) performed by the HERAPDF group.
As in the case of~\cite{Cooper-Sarkar:2016foi} our $\chi^2/DOF \sim 1$ unless combined H1 and ZEUS experimental
data analyzed are cut according to $Q^2 \geq 3.5$ GeV$^2$.

At lower $Q^2$ there is certain disagreement, which is we believe to be explained by the higher-twist (HT)
corrections important in this region. These latter corrections appear to be rather cumbersome
at low $x$~\cite{HT}. Next, as it was shown~\cite{Cvetic1}, it is very promising to use
infrared modifications of the strong coupling constant in our analysis.
Such types of coupling constants modify the low $\mu^2$ behavior of parton densities
and structure functions. What is important, they do not generate additional free parameters.

Following~\cite{Cvetic1}, we are going to use ``frozen'' $a_{\rm fr}(\mu^2)$ \cite{Badelek:1996ap}
and analytic $a_{\rm an}(\mu^2)$ \cite{Shirkov:1997wi} versions
\be
a_{\rm fr}(\mu^2) = a_s(\mu^2 + M^2_{g}),~~~
a_{\rm an}(\mu^2) = a_s(\mu^2) - \frac{1}{\beta_0} \, \frac{\Lambda_{\rm LO}^2}{\mu^2-\Lambda_{\rm LO}^2} \, ,
\label{Ana}
\ee
where $M_{g}$ is a gluon mass with $M_{g}$=1 GeV$^2$ (see~\cite{Badelek:1996ap} and references therein
\footnote{There are a number of various approaches to define the value of this gluon mass and even
the form of its momentum dependence (see, e.g., a recent review~\cite{Deur:2016tte}).}).

As it was shown in~\cite{Kotikov:2017mhk},
the results of the fits carried out when $a_{\rm fr}(\mu^2)$ and $a_{\rm an}(\mu^2)$ are used,
are very similar to the corresponding ones done in~\cite{Kotikov:2012sm}.
Moreover, note that the fits in the cases with ``frozen'' and analytic strong coupling constants look
very much alike (see also~\cite{Cvetic1,KoLiZo}) and describe fairly well the data in the low $Q^2$ region,
as opposed to the fits with a standard coupling constant, which largely fails here.
The results are presented in Table~1 in \cite{Kotikov:2017mhk}):
when the data are cut by $Q^2 \geq $ 1 GeV$^2$,
$\chi^2$ value drops by more than two times. Ditto for the analyses of data with $Q^2 \geq $ 3.5 GeV$^2$ imposed.

\section{Rescaling model}

In the rescaling model~\cite{Close:1984zn} SF $F_2$ and, therefore, valence part of quark densities,
gets modified in the case of a nucleus $A$ at intermediate and large $x$ values $(0.2 \leq x \leq 0.8)$ as follows
\begin{equation}
  F_2^A(x,\mu^2) =
  F_2(x,\mu^2_{A,v}),~~~
  f_{NS}^A(x,\mu^2) =
  f_{NS}(x,\mu^2_{A,v}),
  \label{va.1}
\end{equation}
where a new scale $\mu^2_{A,v}$ is related with $\mu^2$ as
\begin{equation}
  \mu^2_{A,v} = \xi^A_v(\mu^2)\mu^2,~~~
   \xi^A_v(\mu^2) = \Bigl(\lambda_A^2/\lambda_N^2\Bigr)^{a_s(\tilde{\mu}^2)/a_s(\mu^2)} \,
  \label{va.1a}
\end{equation}
where some additional scale $\tilde{\mu}^2=0.66$ GeV$^2$, which was in its turn an initial point
in a $\mu^2$-evolution performed in~\cite{Close:1984zn}; it is then estimated in Appendix~A of that paper.
The quantity $\lambda_A/\lambda_N$ stands for the ratio of quark confinement radii in a nucleus $A$ and nucleon.
The values of $\lambda_A/\lambda_N$ and $\xi^A_v(\mu^2)$ at $\mu^2=20$ GeV$^2$
were evaluated for different nuclei and presented in Tables I and II in~\cite{Close:1984zn}.

Since the factor $ \xi^A_v(\mu^2)$ is $\mu^2$ dependent, it is convenient to transform it
to some $\mu^2$ independent one. To this end, we considered in \cite{Kotikov:2017mhk} the variable
$\ln(\mu^2_{A,v}/\Lambda^2)$, which has the following form (from Eq.~(\ref{va.1a}))
\begin{equation}
\ln\left(\frac{\mu^2_{A,v}}{\Lambda^2}\right) = \ln\left(\frac{\mu^2}{\Lambda^2}\right) \cdot \Bigl(1+ \delta^A_v\Bigr),~~~
\delta^A_v = \frac{1}{\ln\left(\tilde{\mu}^2/\Lambda^2\right)} \,
  \ln\left(\frac{\lambda_A^2}{\lambda_N^2}\right)\,
 \,,
\label{delta}
\end{equation}
where the nuclear correction factor $\delta^A_v$  becomes $\mu^2$ independent. Moreover,
it is seen that two parameters, namely, the scale $\tilde{\mu}$ and ratio $\lambda_A/\lambda_N$,
are combined to form a $\mu^2$-independent quantity.
Using Eqs.~(\ref{va.1}) and/or~(\ref{delta}), we recovered results for $\delta^A_v$, which
can be found in Table 2 in~\cite{Kotikov:2017mhk}.

Since our parton densities contain the variable $s$ defined in Eq.~(\ref{intro:1a}),
it is convenient to consider its $A$ modification. It has the following simple form:
\begin{equation}
s^A_v \equiv \ln \left(\frac{\ln\left(\mu^2_{A,v}/\Lambda^2\right)}{\ln\left(\mu^2_{0}/\Lambda^2\right)}\right)
= s +\ln\Bigl(1+\delta^A_v\Bigr) \approx s +\delta^A_v,~~~
\label{sA}
\end{equation}
i.e. the nuclear modification of the basic variable $s$ depends on the
$\mu^2$ independent parameter $\delta^A_v$, which possesses very small values (see Table 2 in~\cite{Kotikov:2017mhk}).

\section{Rescaling model al low $x$}

The standard evidence coming from earlier studies is that the rescaling model
is inapplicable at small $x$ values (see, for example,~\cite{Efremov:1986mt}).
It looks like it can be related with some simplifications of low $x$ analyses (see, for example,~\cite{Kotikov:1988aa},
where the rise in EMC ratio was wrongly predicted at small $x$ values).

Using an accurate study of DGLAP equations at low $x$ within the framework of the generalized DAS approach,
it is possible to achieve nice agreement with
the experimental data for the DIS structure functon $F_2$ (see~\cite{Kotikov:2017mhk}) and the previous section)\footnote{Moreover,
using an analogous approach, good agreement was also found with the corresponding data for jet multiplicites~\cite{Bolzoni:2012ii}.}.
Therefore, in~\cite{Kotikov:2017mhk} it is suggested that all these indicate toward success in describing the EMC ratio by using the same approach.

We note that the main difference between global fits and DAS approach is in the restriction of
applicability of the latter by low $x$ region only, while the advantage of the DAS approach is in the analytic solution to DGLAP equations.
Thus, in~\cite{Kotikov:2017mhk} we applied the DAS approach to low $x$ region of EMC effect using a simple fact that the rise of parton
densities increases with increasing $Q^2$ values. This way, with scales of PDF evolutions
less than $Q^2$ (i.e. $\mu^2 \leq Q^2$) in nuclear cases, we can directly reproduce the shadowing effect which is observed
in the global fits. Since there are two components (see Eq.~(\ref{8.02})) for each parton density, we have two free parameters
$\mu_{\pm}$ to be fit in the analyses of experimental data for EMC effect at low $x$ values.

An application of the rescaling model at low $x$ can be incorporated at LO as follows:
\begin{eqnarray}
F^A_2(x,\mu^2) &=& e \, f^A_q(x,\mu^2),~~ F^N_2(x,\mu^2) = e \, f_q(x,\mu^2),
\nonumber \\
  f^A_a(x,\mu^2) &=&
  f_a^{A,+}(x,\mu^2) + f_a^{A,-}(x,\mu^2),~~
  f^{A,\pm}_a(x,\mu^2) =
  f^{\pm}_a(x,\mu^2_{A,\pm}) \, .
	\label{8.02A}
\end{eqnarray}
with a similar definition of $\mu^2_{A,\pm}$ as in the previous section (up to replacement
$v \to \pm$). The expressions for $f^{\pm}_a(x,\mu^2)$ are given in Eq.~(\ref{8.02}).
Then, the corresponding values of $s^A_{\pm}$ are found to be
\begin{equation}
  s^A_{\pm} \equiv \ln \left(\frac{\ln\left(\mu^2_{A,\pm}/\Lambda^2\right)}{\ln\left(\mu^2_{0}/\Lambda^2\right)}\right)
  = s +\ln\Bigl(1+\delta^A_{\pm}\Bigr)\, ,
\label{sApm}
\end{equation}
because of the saturation at low $x$ values for all considered $Q^2$ values, which in our case should be related
with decreasing the arguments of ``$\pm$'' component. Therefore, the values of $\delta^A_{\pm}$ should be negative.

\section{Analysis of the low $x$ data for nucleus}

Note that it is usually convenient to study the following ratio
(see Fig.~1 in~\cite{Kulagin:2016fzf})
\begin{equation}
R^{AD}_{F2}(x,\mu^2) = \frac{F^A_2(x,\mu^2)}{F^D_2(x,\mu^2)}\,.
\label{AD}
\end{equation}

Using the fact that the nuclear effect in a deutron is very small (see Table~1 for the values
of $\delta^A_{v}$ and discussions in~\cite{Kulagin:2016fzf})
\footnote{The study of nuclear effects in a deutron can be found in the recent paper~\cite{AKP},
  which also contains short reviews of preliminary investigations.},
we can suggest that
\begin{eqnarray}
  F^D_2(x,\mu^2) &=& e \, f_q(x,\mu^2),~~
  F^A_2(x,\mu^2) =
  e \,
  f^{AD}_q(x,\mu^2),
    \nonumber \\
f^{AD}_a(x,\mu^2) &=&
f_a^{AD,+}(x,\mu^2) + f_a^{AD,-}(x,\mu^2),~~
f^{AD,\pm}_a(x,\mu^2) =
f^{\pm}_a(x,\mu^2_{AD,\pm}) \, .
\label{AD1}
\end{eqnarray}
The expressions for $f^{\pm}_a(x,\mu^2)$ are given in Eq.~~(\ref{8.02}) and
the corresponding values of $s^{AD}_{\pm}$ are found to be
\be
s^{AD}_{\pm} \equiv \ln \left(\frac{\ln\left(\mu^2_{AD,\pm}/\Lambda^2\right)}{\ln\left(\mu^2_{0}/\Lambda^2\right)}\right)
  = s +\ln\Bigl(1+\delta^{AD}_{\pm}\Bigr)\, .
\label{AD2}
\ee

\section{ $A$ dependence  at low $x$ }

Taking NMC experimental data~\cite{Arneodo:1995cs} along with E665 and HERMES Collaborations~\cite{Adams:1995is}
for the EMC ratio at low $x$ in the case of different nuclei, we can find the $A$ dependence of $\delta^{AD}_{\pm}$,
which can be parameterized as follows
\be
- \delta^{AD}_{\pm} = c^{(1)}_{\pm} + c^{(2)}_{\pm} A^{1/3}.
\label{AD2}
\ee

As it was already mentioned in Sec.~2, usage of the analytic coupling constant leads
to the fits with smaller $\chi^2$ values. For example, the values of $c^{(1)}_{\pm}$ and $c^{(2)}_{\pm}$ found
in the combined fit of the data (76 points) when the analytic coupling constant is used (with $\chi^2=89$)
look like
\be
c^{(1)}_{+,an} = -0.055 \pm 0.015,~~ c^{(2)}_{+,an} = 0.068 \pm 0.006,~~
c^{(1)}_{-,an} = 0.071 \pm 0.101,~~ c^{(2)}_{-,an} = 0.120\pm 0.039 \, .
\label{AD2.an}
\ee

Now, using the $A$ dependence (\ref{AD2}), $R^{AD}_{F2}(x,\mu^2)$ values for any nucleus $A$ can be predicted.
What is more, we can consider also the ratios $R^{AD}_{a}(x,\mu^2)$
of parton densities in a nucleus and deutron themselves,
\begin{equation}
R^{AD}_{a}(x,\mu^2) = \frac{f^{AD}_a(x,\mu^2)}{f_a(x,\mu^2)},~~ (a=q,g) \, ,
\label{ADa}
\end{equation}
with $f^{AD}_a(x,\mu^2)$ and ${f_a(x,\mu^2)}$ defined in Eqs.~(\ref{AD1})-(\ref{AD2}) and~(\ref{8.02})-(\ref{intro:1b}), respecively.


Indeed, at LO $R^{AD}_{q}(x,\mu^2)=R^{AD}_{F2}(x,\mu^2)$;  therefore, results for $R^{AD}_{q}(x,\mu^2)$ are already known.
Since all the parameters of PDFs found within the framework of the
gDAS approach are now fixed
we can predict the ratio  $R^{AD}_{g}(x,\mu^2)$ of the gluon densities in a nucleus and nucleon given
in Eqs.~(\ref{8.02}) and (\ref{AD1}), which is currently under intensive studies
(see a recent paper~\cite{Frankfurt:2016qca} and review~\cite{Armesto:2006ph} along with references and discussion therein).

\begin{figure}[t]
\centering
\vskip 0.5cm
\includegraphics[height=0.45\textheight,width=0.8\hsize]{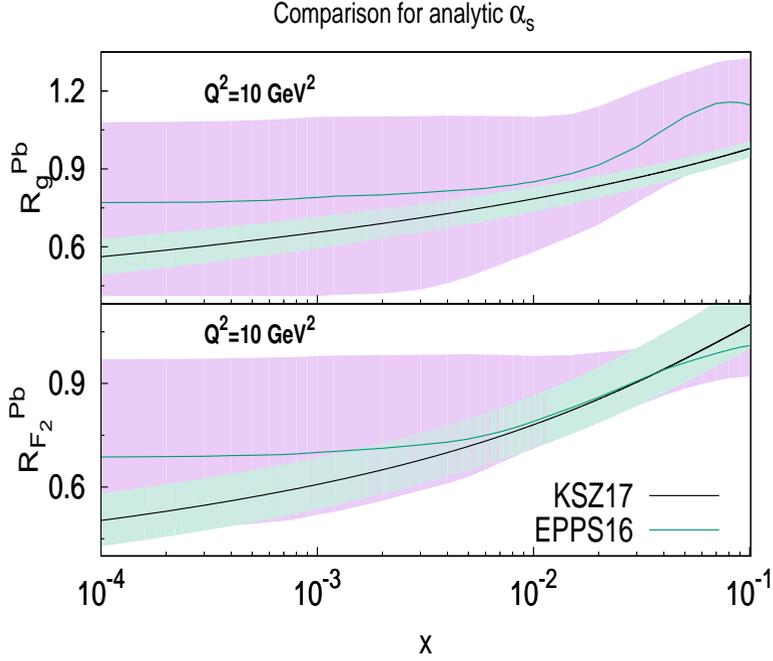}
\vskip -0.3cm
\caption{$x$ dependence of $R^{AD}_{F2}(x,\mu^2)$ and $R^{AD}_{g}(x,\mu^2)$ at $\mu^2$=10 GeV$^2$ for lead data.
A green line with pink band (shows 90$\%$ uncertainties) is taken from the second paper of~\cite{Armesto:2006ph},
while a black one with light green band is obtained in \cite{Kotikov:2017mhk}.}
\end{figure}

The results for $R^{AD}_{F2}(x,\mu^2)$ and $R^{AD}_{g}(x,\mu^2)$, depicted in Fig.~1, show some
difference between these ratios. It is also seen that the difference is similar to that
obtained in a recent EPPS16 analysis~\cite{Eskola:2016oht}
\footnote{
Note that the result for $R^{AD}_{g}(x,\mu^2)$ along with its uncertainty is completely determined
by both the rescaling model and the analytic form for parton densities at low $x$ values we've used.
Therefore, it is clear that the light green band for $R^{AD}_{g}(x,\mu^2)$ should become broader
due to a freedom in using various models.
Also note that a comparison between two uncertainty bands shown in Fig.~3 is in some sense misleading.
The pink band is much broader since the EPPS16 global analysis included a fit to all available data
across quite a wide range in $x$ as opposed to small $x$ consideration adopted in~\cite{Kotikov:2017mhk}.}.
However, what for $R^{AD}_{F2}(x,\mu^2)$ and $R^{AD}_{g}(x,\mu^2)$ themselves (irrespective of other results),
we obtained in \cite{Kotikov:2017mhk} a bit stronger effect at lowest $x$ values, which does in fact not contradict
the experimental data collected by the LHCb experiment (see recent review in~\cite{Winn:2017kwv}). Such a strong
effect is also well compatible with the leading order EPPS09 analysis (which can also be found in~\cite{Winn:2017kwv}).
It will be interesting to delve into more in-depth studies of the ratio $R^{AD}_{g}(x,\mu^2)$, which is one of our aims in the future.





\section{Conclusion}

Using a recent progress in the application of double logarithmic approximations~(see \cite{Q2evo,Kotikov:2012sm}
and~\cite{Bolzoni:2012ii}) to the studies of small $x$ behavior of the structure and fragmentation functions,
respectively, we applied in~\cite{Kotikov:2017mhk} the gDAS approach~\cite{Munich,Q2evo} to examine an EMC $F_2$ structure function ratio
between various nuclei and a deutron. Within a framework of the rescaling model~\cite{Close:1984zn,Close:1983tn}
good agreement between theoretical predictions and respective experimental data was achieved.

The theoretical formul\ae ~contain certain parameters, whose values were fit in
the analyses of experimental data. Once the fits were carried out we had predictions for
the corresponding ratios of parton densities without free parameters. These results were used in \cite{Kotikov:2017mhk} to
predict small $x$ behavior of the gluon density in nuclei, which is at present poorly known.

The ratios $R^{AD}_{a}(x,\mu^2)$ $(a=q,g)$ predicted in  \cite{Kotikov:2017mhk}
are compatible with those
given by various groups working in the area. From our point of view, it is quite valuable
that the application of the rescaling model~\cite{Close:1984zn,Close:1983tn} provided us with
very simple forms for these ratios.
It should also be mentioned that without any free parameters we also predicted the ratio
$R^{AD}_{c}(x,\mu^2)$ of charm parts, $F^{A}_{2c}(x,\mu^2)$ and $F^{D}_{2c}(x,\mu^2)$, of the respective structure functions.
This latter ratio has a simple form and it is very similar \cite{Kotikov:2017mhk}
to the corresponding ratio of the complete structure functions $F^{A}_{2}(x,\mu^2)$ and $F^{D}_{2}(x,\mu^2)$.
We hope that the results for $R^{AD}_{c}(x,\mu^2)$ can be compared with future
experimental data obtained at  Electron-Ion Collider (see~\cite{Chudakov:2016ytj}).

Concluding, we would like to note that an extension of the results obtained in~\cite{Kotikov:2017mhk}
to the range $x\sim 0.1$, with the valence quark density (see its parametrization in~\cite{Illarionov:2010gy}) taken into account,
leads to the clear antishadowing effect in agreement with other studies~\cite{Kotikov:2018cju}.\\


Support by the National Natural Science Foundation of China (Grant
No. 11575254) is acknowledged. A.V.K. and B.G.S. thank Institute of Modern Physics
for invitation. A.V.K. is also grateful to the CAS President's International Fellowship Initiative
(Grant No.~2017VMA0040) for support.
The work of A.V.K. and B.G.S. was in part supported by the RFBR Foundation through the Grant No.~16-02-00790-a.

\end{document}